\documentclass[aip,jmp, eqsecnum,amsmath,amssymb,preprint, reprint]{revtex4-1}
\usepackage{graphicx}
\usepackage{dcolumn}
\usepackage{bm}
\usepackage[utf8]{inputenc}
\usepackage[T1]{fontenc}
\usepackage{mathptmx}
\usepackage{multirow}
\begin{document}
\title[]{Geometric phases for finite-dimensional systems -- the roles of Bargmann\\ Invariants, Null Phase Curves and the Schwinger 
Majorana SU(2) framework}

\author{K. S. Akhilesh}
\email{akhi.s.karagadde@gmail.com}
\affiliation{Department of Studies in Physics, University of Mysore, Manasagangotri, Mysuru 570006, India}
\author{Arvind}
\email{arvind@iiser.mohali.ac.in}
\affiliation{Department of Physical Sciences, Indian Institute of Science Education and Research (IISER) Mohali, Sector 81 SAS Nagar, Manauli PO 140306, 
Punjab, India}

\author{S. Chaturvedi}
\email{subhash@iiserb.ac.in}
\affiliation{Department of Physics, Indian Institute of Science Education and Research (IISER) Bhopal, Bhopal Bypass Road, Bhauri, Bhopal 462066, India}

\author{K. S. Mallesh}
\email{ksmallesh@gmail.com}
\affiliation{Department of Studies in Physics, University of Mysore, Manasagangotri, Mysuru 570006, India}
\affiliation{Regional Institute of Education (NCERT), Manasagangotri, Mysuru 570006, India}

\author{N. Mukunda}
 \email{nmukunda@gmail.com}
\affiliation{INSA Distinguished Professor, Indian Academy of Sciences, C V Raman Avenue, Sadashivanagar, Bangalore 560080, India}

\date{\today}
\begin{abstract}
We present a study of the properties of Bargmann Invariants (BI) and
Null Phase Curves (NPC) in the theory of the geometric phase for
finite dimensional systems. A recent suggestion to exploit the
Majorana theorem on symmetric SU(2) multispinors is combined with
the Schwinger oscillator operator construction to develop efficient
operator based methods to handle these problems. The BI is described
using intrinsic unitary invariant angle parameters, whose algebraic
properties as functions of Hilbert space dimension are analysed
using elegant group theoretic methods. The BI-geometric phase
connection, extended by the use of NPC's, is explored in detail, and
interesting new experiments in this subject are pointed out.
\end{abstract}

\maketitle

The geometric phase was originally discovered in a quantum
mechanical context based on several physically reasonable
assumptions \cite{1}. Thereafter our understanding of it has evolved in
several steps to a level where many of the original assumptions have
been shown to be unnecessary \cite{2}. In the kinematic approach the
minimal conditions needed to be able to define the geometric phase
have been identified \cite{3}. In the process the role of the Bargmann
Invariants, and later of the so-called Null Phase Curves as basic
ingredients of the theory, have been clarified \cite{4}. The original
connection between Bargmann Invariants and geometric phases based on
geodesics in quantum mechanical ray spaces has been greatly enlarged
by showing that the geodesics can be replaced by the much more
numerous Null Phase Curves.

Bargmann Invariants (BI) are of various integral orders, the lowest
nontrivial one being of order three. The fourth and higher order
BI's can in principle be reduced to those of third order, which is
thus the primitive one. It is therefore natural to study these in
some detail. It has recently been pointed out that for finite
dimensional systems the general properties of these BI's are not
much known, especially for high dimensions \cite{5}. It has then been
shown that the calculation of these BI's can be handled in a uniform
and efficient manner using Majorana's theorem concerning symmetric
SU(2) multispinors \cite{6}. The calculations go back to computing solid
angles on Poincar\'{e} spheres familiar from polarization optics.

The purpose of the present work is to explore the properties and
parametrisations of the third order Bargmann Invariant. We show  how
its intrinsic invariance and other algebraic properties can be
brought out elegantly using group theoretical methods. In particular
it can be parametrized in a natural and intrinsic manner using
unitary invariant angle variables. We bring out the dependence of
the algebraic properties of these variables on the state space
dimension. We then combine the Majorana theorem with the Schwinger
oscillator construction of SU(2) representations \cite{7}, and study both
Bargmann Invariants and Null Phase Curves in this new framework.

The contents of this paper are organised as follows. Section I
recalls definitions of BI's and geometric phases, and the connection
between them based on ray space geodesics. It also introduces a
natural set of six angle parameters associated with any third order
BI, which are invariant under all unitary transformations. Section II
examines the extent to which these angles are algebraically
independent. Using group theoretic methods, it is shown that while
for two-dimensional systems only five of the six angles are
independent, for dimensions three and higher all six are
independent. Interestingly it is shown that for the subset of
coherent states of a one-dimensional oscillator, only five of the
six angles are independent. Section III describes briefly the family
of Null Phase Curves (NPC) in Hilbert and ray spaces which have been
shown  to be basic to geometric phase theory. The definition,
important properties and procedure for construction of NPC's are
given in a concise manner. Section IV studies both BI's and NPC's for
finite dimensional systems using the Schwinger--Majorana framework.
This is a combination of the Schwinger oscillator treatment of
quantum angular momentum theory, and the Majorana theorem on
symmetric SU(2) multispinors. Section V contains some Concluding
Remarks, and the Appendix presents the basic features of the
Schwinger--Majorana framework which allows a uniform description of
all finite dimensional Hilbert spaces.

\section{Three-vertex Bargmann Invariants -- invariances, intrinsic parameters, connection to geometric phases}

Let $\mathcal{H}$ be a complex Hilbert space, of finite or infinite
dimension, pertaining to some quantum system. Vectors in
$\mathcal{H}$ are $\psi, \phi, \ldots$, and the inner product is
$(\phi, \psi)$ or $\langle \phi|\psi\rangle$. The unit sphere
$\mathcal{B}\subset \mathcal{H}$, and the space $\mathcal{R}$ of
unit rays, are
\begin{eqnarray}
\mathcal{B}&=&\{\psi\in \mathcal{H}|(\psi, \psi)=1\}\subset \mathcal{H},\nonumber\\
\mathcal{R}&=&\{\rho(\psi)=|\psi\rangle\langle \psi|\ |\ \psi\in
\mathcal{B}\}.
\end{eqnarray}

Neither $\mathcal{B}$ nor $\mathcal{R}$ is a linear space, they are
related by a projection map $\pi$:
\begin{equation}
\pi: \mathcal{B}\rightarrow \mathcal{R}:\quad \psi\rightarrow
\rho(\psi).
\end{equation}
Thus $\mathcal{B}$ is a U(1) bundle over $\mathcal{R}$ as base.

If the complex dimension of $\mathcal{H}$ is finite, $n$ say, then
$\mathcal{B}$ and $\mathcal{R}$ are spaces of real dimensions
$(2n-1), 2(n-1)$ respectively.

Let $\psi_j, j=1, 2, 3$  be any three vectors in $\mathcal{B}$,
pairwise linearly independent and nonorthogonal. They define the
third order or three-vertex Bargmann invariant (BI)
\begin{equation}
\Delta_3(\psi_1, \psi_2, \psi_3)=(\psi_1, \psi_2)( \psi_2,
\psi_3)(\psi_3, \psi_1),
\end{equation}
which is basic to geometric phase theory. This expression is
nonzero, in general complex, and is actually defined on
$\mathcal{R}$ since
\begin{equation}
\Delta_3(\psi_1, \psi_2, \psi_3)
=Tr(\rho(\psi_1)\rho(\psi_2)\rho(\psi_3)). \end{equation} The
relation to geometric phases, as originally established, arises as
follows \cite{8}. For any two `nonorthogonal' points $\rho(\psi_1),
\rho(\psi_2)$ in $\mathcal{R}$, there is a (unique shorter) geodesic
(with respect to the Fubini-Study metric on $\mathcal{R}$)
connecting them. Choose $\psi_1, \psi_2$ to be `in phase' in the
Pancharatnam sense \cite{9}:
\begin{equation}
(\psi_1, \psi_2)=\cos\frac{1}{2}\theta_0, \quad 0<\theta_0<\pi.
\label{2.5}
\end{equation}
Then the geodesic $C_{{\rm geo}, 12}$ from $\rho(\psi_1)$ to
$\rho(\psi_2)$ is the projection by $\pi$ of a parametrised curve
$\mathcal{C}_{{\rm geo}, 12}$ in $\mathcal{B}$ from $\psi_1$ to
$\psi_2$:
\begin{eqnarray}
C_{{\rm geo}, 12}&=&\pi[\mathcal{C}_{{\rm geo}, 12}],\nonumber\\
\mathcal{C}_{{\rm geo},
12}&=&\{\psi(s)=\frac{1}{\sin\frac{1}{2}\theta_0}(\psi_1\sin\frac{1}{2}(1-s)\theta_0+
\psi_2\sin\frac{1}{2}s\theta_0)|\nonumber\\&& 0\le s\le 1\}\subset
\mathcal{B}.
\label{2.6}
\end{eqnarray}
Now given the vertices $\rho(\psi_1), \rho(\psi_2), \rho(\psi_3)$ of
$\Delta_3(\psi_1, \psi_2, \psi_3)$, connect them by successive
geodesics $ C_{{\rm geo}, 12}, C_{{\rm geo}, 23}$ and $ C_{{\rm
geo}, 31}$. (Of course we cannot choose $\psi_1, \psi_2, \psi_3$ to
all obey conditions like $(\ref{2.5})$ in general!) Then their union
\begin{eqnarray}
&& C:\rho(\psi_1)\rightarrow \rho(\psi_1):\nonumber\\
&& C= C_{{\rm geo}, 12} \cup C_{{\rm geo}, 23}\cup C_{{\rm geo}, 31}
\end{eqnarray}
gives a geodesic triangle in $\mathcal{R}$, a closed cyclic
evolution in the quantum mechanical state space. This evolution can
be produced by a suitable (time dependent) Hamiltonian operator via
the Schr\"odinger equation, and the corresponding geometric phase
turns out to be
\begin{equation}
\varphi_{\rm geom}[ C_{{\rm geo}, 12} \cup C_{{\rm geo}, 23}\cup
C_{{\rm geo}, 31}]=-\arg \Delta_3(\psi_1, \psi_2, \psi_3).
\label{2.8}
\end{equation}
Actually, the use of a Hamiltonian and Schr\"odinger equation are not
essential; this connection between geometric phases and BI's can be
understood more directly in the kinematic approach. We return to
this and to its nontrivial generalisation in the sequel.

At this point we turn to a study of the invariances and intrinsic
properties of the BI $\Delta_3(\psi_1, \psi_2, \psi_3)$. If $U$ is
any unitary transformation on $\mathcal{H}$, we have the obvious
invariance property:
\begin{equation}
\psi'_j=U\psi_j, j=1, 2, 3: \Delta_3(\psi'_1, \psi'_2, \psi'_3)=
\Delta_3(\psi_1, \psi_2, \psi_3).
\label{2.9}
\end{equation}
The BI itself can be parametrised by six intrinsic angle type
variables, two for each factor:
\begin{eqnarray}
jk=12, 23, 31: (\psi_j, \psi_k)=e^{i\varphi_{jk}}\cos \frac{1}{2}\theta_{jk},\nonumber\\
0<\theta_{jk}<\pi, \quad 0\le \varphi_{jk}<2\pi.
\label{2.10}
\end{eqnarray} The
six variables $\theta_{jk}, \varphi_{jk}$ are all defined at the
level of $\mathcal{B}$, and are of course invariant under the
transformations $(\ref{2.9})$, i.e., they are unitary invariants. They are
intrinsic to the triad $\{\psi_j\}$. If we make independent phase
changes in the $\psi_j$, the $\theta_{jk}$ are unchanged while the
$\varphi_{jk}$ change in a simple way:
\begin{equation}\psi'_j=e^{i\alpha_j}\psi_j:\quad \varphi'_{jk}=\varphi_{jk}-\alpha_j+\alpha_k.
\label{2.11}
\end{equation}
Thus the individual $\varphi_{jk}$ are not ray space quantities but
their sum is invariant under $(\ref{2.11})$, and according to $(\ref{2.8})$ is a
geometric phase:
\begin{equation}
-\arg \Delta_3(\psi_1, \psi_2, \psi_3)= -(\varphi_{12}+
\varphi_{23}+ \varphi_{31})= \hbox{a geometric phase}.
\label{2.12}
\end{equation}
We can see that the six angles $\theta_{jk}, \varphi_{jk}$ determine
the triad $\{\psi_j\}$, equivalently their configuration, upto an
overall unitary transformation $U$ on $\mathcal{H}$. The important
question is the extent to which they are algebraically independent.

\section{Algebraic independence of Bargmann Invariant parameters in two and three dimensions, the case of coherent states}

\setcounter{equation}{0}

The question we wish to answer is this: can the values of
$\theta_{jk}, \varphi_{jk}$ be chosen independently, and will they
then lead to a definite triad of vectors $\{\psi_j\}$ upto an
overall unitary transformation? We will find that the answer depends
on the dimension of $\mathcal{H}$, and also on whether any overall
restrictions are placed on the possible choices of the $\psi_j$.

\subsection*{The case $\dim \mathcal{H}=n=2$}
The full unitary group on $\mathcal{H}$ is the four-parameter U(2),
and the $\psi_j's$ can be represented by two component complex
column vectors in an unspecified orthonormal basis (ONB). In any
case, $\psi_3$, say, can be expressed as a linear combination of
$\psi_1$ and $\psi_2$ as the latter are linearly independent, with
coefficients determined by $\theta_{jk}, \varphi_{jk}$:
\begin{eqnarray}
\psi_3&=&\frac{1}{S_{12}^2}\{e^{-i\varphi_{31}} (C_{31} - e^{-i\varphi_g}C_{12}C_{23})\psi_1 + \nonumber\\
&&~~~~~~~~~~~~~~~~~~~~~~~~e^{i\varphi_{23}} (C_{23} - e^{i\varphi_g}C_{31}C_{12})\psi_2\}, \nonumber\\
\varphi_g&=&-(\varphi_{12}+\varphi_{23}+\varphi_{31}),\nonumber\\
S_{12}&=&\sin\frac{1}{2}\theta_{12}, \quad
C_{12}=\cos\frac{1}{2}\theta_{12},\ldots.
\label{3.1}
\end{eqnarray}
Such a relation need not hold in higher dimensions.

To proceed further we use a group theoretic approach. Using the
freedom of U(2) action we can assume without loss of generality
that in some ONB
\begin{equation}
\psi_1=\left(\begin{array}{c}1\\ 0\end{array}\right).
\label{3.2}
\end{equation}
The stability group of this $\psi_1$ is a one-dimensional U(1)
subgroup $\text{H}_1 \subset $ U(2):
\begin{equation}
\text{H}_1=\left\{\left(\begin{array}{cc}1 & 0\\ 0&
e^{i\beta}\end{array}\right), 0\le \beta<2\pi\right\}\subset \text{U(2)}.
\label{3.3}
\end{equation}
For independently given angles $\theta_{12}, \varphi_{12}$ we can
use $\text{H}_1$ action to achieve
\begin{equation}
\psi_2=e^{i\varphi_{12}}\left(\begin{array}{c}C_{12}\\
S_{12}\end{array}\right),
\label{3.4}
\end{equation}
after which there is no more U(2) action possible. Then for
independently given angles $\theta_{31}, \varphi_{31}$ the vector
$\psi_3$ is necessarily of the form
\begin{equation}
\psi_3=e^{-i\varphi_{31}}\left(\begin{array}{c}C_{31}\\
e^{i\phi}S_{31}\end{array}\right),\quad 0\le \phi< 2\pi,
\label{3.5}
\end{equation}
bringing in one additional angle $\phi$ independent of $\theta_{12},
\theta_{31}, \varphi_{12}, \varphi_{31}$. Thus a given triad
$\{\psi_j\}$ leads to exactly five independent U(2) invariant angles
$\theta_{12}, \theta_{31}, \varphi_{12}, \varphi_{31}, \phi$. The
two remaining U(2) invariant angles $\theta_{23}, \varphi_{23}$ are
to be found from
\begin{eqnarray}
&& (\psi_2, \psi_3)= e^{i\varphi_{23}}C_{23}=e^{-i(\varphi_{12}+\varphi_{31})}(C_{12}C_{31}+e^{i\phi}S_{12}S_{31}),\nonumber\\
&& i.e., \quad C_{23}=e^{i\varphi_g}(C_{12}C_{31}
+e^{i\phi}S_{12}S_{31}).
\label{3.6}
\end{eqnarray}
Since
\begin{equation}
|C_{12}C_{31}+e^{i\phi}S_{12}S_{31}|\le  C_{12}C_{31}+
S_{12}S_{31}=\cos\frac{1}{2}( \theta_{12}- \theta_{31})\le 1,
\end{equation}
we see that $\theta_{23}$ and $\varphi_g$ are unambiguously
determined by eq. $(\ref{3.6})$ in terms of $\theta_{12}, \theta_{31}, \phi$
and so cannot be independently chosen. Once $\varphi_g$ has been
determined, $\varphi_{23}$ follows from
\begin{equation}
\varphi_{23}=-(\varphi_g+ \varphi_{12}+ \varphi_{31}).
\label{3.8}
\end{equation}

We thus find that five algebraically independent intrinsic U(2)
invariant angles can be chosen in several equivalent ways:
$\theta_{12}, \theta_{31},  \phi, \varphi_{12}, \varphi_{31}$ or
$\theta_{12}, \theta_{31},  \theta_{23}, \varphi_{12}, \varphi_{31}$
or $\theta_{12}, \theta_{31},  \varphi_{23}, \varphi_{12},
\varphi_{31}$. (Other choices are also easily found). In any case,
the six U(2) invariant angles $\theta_{jk}, \varphi_{jk}$ are not
algebraically independent. Each of the three choices of
algebraically independent angles listed above is of course at the
level of the space $\mathcal{B}$. Taking for example the second set
consisting of $\theta_{12}, \theta_{31}, \theta_{23}, \varphi_{12},
\varphi_{31}$, we see that while the $\theta_{jk}$ `descend' to the
ray space $\mathcal{R}$, neither $\varphi_{12}$ nor $\varphi_{31}$
do so since under independent phase changes $(\ref{2.11})$ in the vectors
$\psi_j$ they are not preserved but change independently. Thus a
triad of vertices $\rho(\psi_j), j=1, 2, 3$ in $\mathcal{R}$,
determining a geodesic triangle in $\mathcal{R}$, is determined
intrinsically by exactly three algebraically independent angles
$\theta_{jk}$. For dim $\mathcal{H}=2, \mathcal{R}$ is the
(Poincar\'{e}) sphere $S^2$, and as is well known a geodesic
triangle on $S^2$ is intrinsically determined by three independent
angle parameters. It was for the cyclic evolution of the
polarization state of plane electromagnetic waves around such a
triangle on $S^2$ that the geometric phase was found in
Pancharatnam's pioneering work \cite{9}.

\subsection*{The case of one-dimensional coherent states}

It is interesting that a remarkably similar situation occurs in the
case of the infinite dimensional Hilbert space
$\mathcal{H}=L^2(\mathbb{R})$, in the context of the coherent
states. This Hilbert space carries an irreducible representation of
the Heisenberg canonical commutation relation for hermitian
operators $\hat{q}, \hat{p}$ or their nonhermitian combinations
$\hat{a}, \hat{a}^\dag$:
\begin{eqnarray}
&& [\hat{q}, \hat{p}]= i\ . \ 1;\nonumber\\
&& \hat{a}=\frac{1}{\sqrt{2}} (\hat{q}+i\hat{p}),
\hat{a}^\dag=\frac{1}{\sqrt{2}} (\hat{q}-i\hat{p}): [\hat{a},
\hat{a}^\dag]=1.
\end{eqnarray}
The eigenstates of $\hat{a}^\dag\hat{a}$ (Fock states), and the
right eigenstates of $\hat{a}$ (coherent states), are related to one
another as follows:
\begin{eqnarray}
&& n=0, 1, 2, \ldots: |n\rangle = \frac{(\hat{a}^\dag)^n}{\sqrt{n!}}|0\rangle, \quad \hat{a}^\dag \hat{a}|n\rangle=n|n\rangle;\nonumber\\
&& z\in \mathbb{C}: |z\rangle =
e^{-\frac{1}{2}|z|^2}\sum_{n=0}^{\infty}
\frac{z^n}{\sqrt{n}!}|n\rangle,\quad \hat{a}|z\rangle=z|z\rangle.
\end{eqnarray}
The former give an ONB in $\mathcal{H}$, the latter are
nonorthogonal and overcomplete:

\begin{eqnarray}
&& \langle n'|n\rangle =\delta_{n', n}, \qquad \sum_{n=0}^{\infty}|n\rangle\langle n|=1;\nonumber\\
&& \langle z'|z\rangle=e^{-\frac{1}{2}|z'-z|^2+i{\rm Im}(z'^*z)},
\int\frac{d^2z}{\pi}|z\rangle\langle z|=1.
\end{eqnarray}
Now define a subset $\mathcal{M}\subset \mathcal{H}$, not a
subspace, by
\begin{equation}
\mathcal{M}=\{e^{i\alpha}|z\rangle |0\le \alpha< 2\pi, z\in
\mathbb{C}\}.
\end{equation}
There is a natural four parameter group $\text{G}_4$ of unitary
transformations on $\mathcal{H}$ which is generated by
$\{\mathbb{I}, \hat{q}, \hat{p}, \hat{a}^\dag\hat{a}\}$ and acts
transitively on $\mathcal{M}$:
\begin{eqnarray}
&& U(\alpha_0, z_0, \theta_0)= e^{i\alpha_0}D(z_0) e^{-i\theta_0 \hat{a}^\dag \hat{a}}\in \text{G}_4, D(z_0)= e^{z_0\hat{a}^\dag-z_0^*\hat{a}}: \nonumber\\
&& U(\alpha_0, z_0, \theta_0) e^{i\alpha}|z\rangle
=e^{i(\alpha+\alpha_0+ {\rm Im}(z_0z^*e^{i\theta_0}))}|z
e^{-i\theta_0}+z_0\rangle.
\end{eqnarray}
(The composition law in $\text{G}_4$ is easily found but not needed for our
purposes).

Let now $|\psi_j\rangle, j=1, 2, 3$ be a triad of vectors with the
restriction that each of them belong to $\mathcal{M}$, and let the
six angles $\theta_{jk}, \varphi_{jk}$ be as in eq. $(\ref{2.10})$. These
angles are of course invariant under $\text{G}_4$ action. Using a suitable
$\text{G}_4$ element we can begin by assuming without loss of generality
\begin{equation}
|\psi_1\rangle=|0\rangle.
\end{equation}
The stability group of this $|\psi_1\rangle$ is the U(1) subgroup
\begin{equation}
{\text H}_1=\{U(0, 0, \theta_0)|0\le \theta_0<2\pi\}\subset \text{G}_4.
\end{equation}
Now for any given independent $\theta_{12}, \varphi_{12}$ we can use
$\text{H}_1$ action to achieve
\begin{eqnarray}
|\psi_2\rangle=e^{i\varphi_{12}}|r\rangle,\nonumber\\
e^{-\frac{1}{2}r^2}=C_{12}, \quad r>0.
\end{eqnarray}
At this point, all of $\text{G}_4$ has been `used up'. Now given any
$\theta_{31}, \varphi_{31}$ independent among themselves and of
$\theta_{12}, \varphi_{12}$, we find that $|\psi_3\rangle$
necessarily has the form
\begin{eqnarray}
|\psi_3\rangle=e^{-i\varphi_{31}}|r'e^{i\phi'}\rangle,\nonumber\\
e^{-\frac{1}{2}r'^2}=C_{31}, \quad 0\le \phi'<2\pi.
\end{eqnarray}
 Thus one additional angle $\phi'$ independent of $\theta_{12}, \theta_{31}, \varphi_{12}, \varphi_{31}$ has appeared (compare with eq. 
 $(\ref{3.5}))$,
so any triad $\{|\psi_j\rangle\}$ of the prescribed type is
characterised by exactly five independent $\text{G}_4$ invariant angles
$\theta_{12}, \theta_{31}, \varphi_{12}, \varphi_{31}, \phi'$. The
remaining two $\text{G}_4$ invariant angles $\theta_{23}, \varphi_{23}$ are
to be found from
\begin{eqnarray}
&& \langle \psi_2|\psi_3\rangle =e^{i\varphi_{23}}C_{23}=e^{-i\varphi_{12}-i\varphi_{31}}\langle r| r'e^{i\phi'}\rangle,\nonumber\\
&& i.e., C_{23}=e^{i\varphi_g}C_{12}C_{31}e^{rr'e^{i\phi'}}\nonumber\\
&& =e^{i(\varphi_g+rr'\sin\phi')} C_{12}C_{31}e^{rr'\cos\phi'}.
\label{3.18}
\end{eqnarray}
Since
\begin{equation}
C_{12}C_{31}e^{rr'\cos\phi'}=e^{-\frac{1}{2}r^2-\frac{1}{2}r'^2+rr'\cos\phi'}\le
e^{-\frac{1}{2}(r-r')^2}\le 1,
\end{equation}
we see that $\theta_{23}$ and $\varphi_g$ are unambiguously
determined by eq. $(\ref{3.18})$ in terms of $\theta_{12}, \theta_{31},
\phi'$. Once $\varphi_g$ has been determined, $\varphi_{23}$ follows
from eq. $(\ref{3.8})$ again.

In this example, then, out of the six angles $\theta_{jk},
\varphi_{jk}$ only five are algebraically independent. As
independent sets we can choose for example $\theta_{12},
\theta_{31}, \phi', \varphi_{12},   \varphi_{31}$ or $\theta_{12},
\theta_{31}, \theta_{23}, \varphi_{12},   \varphi_{31}$, etc. The
overall similarity to the case of $\dim\mathcal{H}=2$ is perhaps
unexpected.

At this point we return to $\mathcal{H}$ of finite dimensions.

\subsection*{The case $\dim \mathcal{H}=n=3$}

Now we find that all six angles $\theta_{jk}, \varphi_{jk}$ are
algebraically independent. Unitary transformations on $\mathcal{H}$
constitute the nine-parameter group U(3); using its action we can
arrange
\begin{equation}
\psi_1=\left(\begin{array}{c}1\\ 0\\ 0\end{array}\right)
\label{3.20}
\end{equation}
in a suitable ONB. The stability group of $\psi_1$ is a U(2)
subgroup in U(3):
\begin{equation}
\text{H}_1=\left\{\left(\begin{array}{ccc}
1& 0& 0\\
0 &\multicolumn{2}{c}{\multirow{2}{*}{$u$}}\\
0 & \multicolumn{2}{c}{}
\end{array}\right)|u\in \text{U}(2) \right\}\subset \text{U}(3).
\end{equation}
For given independent choices of $\theta_{12}, \varphi_{12}$ we use
$\text{H}_1$ action to achieve
\begin{equation}
\psi_2=e^{i\varphi_{12}}\left(\begin{array}{c}C_{12}\\ S_{12}\\
0\end{array}\right).
\label{3.22}
\end{equation}
The stability group for the pair $\psi_1, \psi_2$ is a U(1)
subgroup:
\begin{equation}
\text{H}_2=\left\{\left(\begin{array}{ccc}1& 0& 0\\ 0 & 1 & 0\\ 0 &0 &
e^{i\beta}\end{array}\right)|0\le \beta <2\pi \right\}\subset
\text{H}_1\subset  \text{U}(3).
\end{equation}
Now for given independent choices of $\theta_{31}, \varphi_{31}$ we
use $\text{H}_2$ action to achieve
\begin{equation}
\psi_3=e^{-i\varphi_{31}}\left(\begin{array}{c}C_{31}\\
e^{i\phi}S_{31}\cos\xi\\ S_{31}\sin\xi\end{array}\right),
0<\xi<\pi/2, 0\le \phi<2\pi.
\label{3.24}
\end{equation}
This brings in two additional angles $\phi, \xi$ independent of
$\theta_{12}, \theta_{31}, \varphi_{12}, \varphi_{31}$; and all the
freedom of U(3) action has been `used up'. The triad $\{\psi_j\}$ is
intrinsically characterised by six independent U(3) invariant angles
$\theta_{12}, \theta_{31}, \varphi_{12}, \varphi_{31}, \phi, \xi$.
The two remaining original U(3) invariant angles $\theta_{23},
\varphi_{23}$ are to be found from
\begin{eqnarray}
&& (\psi_2, \psi_3)=e^{i\varphi_{23}}C_{23}\nonumber\\
&& =e^{-i(\varphi_{12} + \varphi_{31})}(C_{12}C_{31}+e^{i\phi}S_{12}S_{31}\cos \xi),\nonumber\\
&& i.e., C_{23}=e^{i\varphi_{g}}(C_{12}C_{31}
+e^{i\phi}S_{12}S_{31}\cos \xi).
\label{3.25}
\end{eqnarray}
In contrast to eq. $(\ref{3.6})$, now (for any given $\theta_{12},
\theta_{31}$) the pair $\theta_{23}, \varphi_{g}$ is unambiguously
determined by the pair $\phi, \xi$ and so they can be independently
specified. We are free to choose $\theta_{12}, \theta_{31},
\varphi_{12}, \varphi_{31}, \phi, \xi$ or $\theta_{jk},
\varphi_{jk}$ as six independent U(3) invariants characterising
$\{\psi_j\}$.

As long as we deal with the three-vertex BI $\Delta_3(\psi_1,
\psi_2, \psi_3)$, the situation for $\dim \mathcal{H}=n\ge 4$ is the
same as for $n=3$, since  $\psi_1, \psi_2, \psi_3$ always lie in
some three-dimensional subspace in $\mathcal{H}$.

\subsection*{Geometric phase formulae for geodesic triangles}

In Hilbert spaces of dimensions 2 and 3 the geometric phase results
of eqs. $(\ref{2.8}, \ref{2.12})$ can be given more explicitly. For the passages
to the ray spaces, $\mathcal{B}\rightarrow \mathcal{R}$, in the two
cases, we use eqs. $(\ref{3.2}, \ref{3.4}, \ref{3.5})$ and eqs. $(\ref{3.20}, \ref{3.22},\ref{3.24})$
respectively, drop overall phase factors, and obtain expressions for
the vertices:
\begin{eqnarray}
&& \rho_j=\psi_j^{(0)} \psi_j^{(0)^+},\quad j=1, 2, 3;\nonumber\\
n=2\quad && \psi_1^{(0)}=
\left(\begin{array}{c}1\\0\end{array}\right), \psi_2^{(0)}=
\left(\begin{array}{c}C_{12}\\S_{12}\end{array}\right),
\psi_3^{(0)}= \left(\begin{array}{c}C_{31}\\e^{i\phi}S_{31}\end{array}\right);\nonumber\\
n=3\quad && \psi_1^{(0)}=
\left(\begin{array}{c}1\\0\\0\end{array}\right),~~~~~~ \psi_2^{(0)}=
\left(\begin{array}{c}C_{12}\\S_{12}\\ 0\end{array}\right),\nonumber\\
&&\psi_3^{(0)}=
\left(\begin{array}{c}C_{31}\\e^{i\phi}S_{31}\cos\xi\\S_{31}\sin
\xi\end{array}\right).
\end{eqnarray}

For $n=2$, a general geodesic triangle on $\mathcal{R}=\mathbb{S}^2$ is
intrinsically characterized by three independent angle parameters
$\theta_{12}, \theta_{31}, \phi$. This is familiar from spherical
trigonometry on $\mathbb{S}^2$. For $n=3$, ray space $\mathcal{R}$ is a
simply connected four dimensional region $\vartheta \subset
\mathbb{S}^7\subset\mathbb{R}^8$, a (small) portion of the unit sphere in
real eight dimensional Euclidean space \cite{10}. A geodesic triangle on
$\vartheta$ is intrinsically characterized by four independent angle
parameters $\theta_{12}, \theta_{31}, \phi, \xi$. From eqs. $(\ref{3.6},
\ref{3.25})$ the geometric phases are:
\begin{eqnarray}
n=2\quad && \varphi_{geom} [\hbox{geodesic triangle on}\ \mathbb{S}^2, \hbox{vertices}\ \rho_1^{(0)}, \rho_2^{(0)}, \rho_3^{(0)}]\nonumber\\
&& =-\arg\ \Delta_3(\psi_1^{(0)}, \psi_2^{(0)}, \psi_3^{(0)})=-(\varphi_{12}+\varphi_{23}+ \varphi_{31})\nonumber\\
&& =-\arg
(1+e^{i\phi}\tan\frac{1}{2}\theta_{12}\tan\frac{1}{2}\theta_{31});\qquad
~~~~~~~~(a)\nonumber\\
n=3\quad && \varphi_{geom} [\hbox{geodesic triangle on}\ \vartheta, \hbox{vertices}\ \rho_1^{(0)}, \rho_2^{(0)}, \rho_3^{(0)}]\nonumber\\
&& =-\arg\ \Delta_3(\psi_1^{(0)}, \psi_2^{(0)}, \psi_3^{(0)})=-(\varphi_{12}+\varphi_{23}+ \varphi_{31})\nonumber\\
&& =-\arg
(1+e^{i\phi}\tan\frac{1}{2}\theta_{12}\tan\frac{1}{2}\theta_{31}\cos\xi).\qquad
(b)\nonumber\\
\label{3.27}
\end{eqnarray}

For $n=2$, this is the original result of Pancharatnam; for $n=3$ we
have a genuine generalisation of Pancharatnam's result \cite{9}, \cite{10}.

\section{Definition, properties, construction of Null Phase Curves, generalized Bargmann Invariant-geometric phase connection}

\setcounter{equation}{0}

It was mentioned in Section I that an extensive generalisation of
the BI-geometric phase connection $(\ref{2.8})$ exists. It turns out that
the ray space geodesics appearing on the left hand side of eq. $(\ref{2.8})$
can each be replaced by a so-called Null Phase Curve (NPC), with no
change on the right hand side. Thus, the (negative of the) phase of
a BI is the geometric phase for many cyclic ray space evolutions,
not just for  the one along the sides of a geodesic triangle.

It is an interesting fact that for $\dim \mathcal{H}=2$, NPC's
coincide with geodesics. However, for $\dim \mathcal{H}\ge 3$, given
any two nonorthogonal points $\rho_1, \rho_2\in \mathcal{R}$, there
exist infinitely many NPC's connecting them, the geodesic being just
one of them. In this Section we define, describe and outline the
construction of the most general NPC connecting any two given
non-orthogonal points in ray space. This is a summary of the results
in references \cite{4}. The spaces $\mathcal{B}, \mathcal{R}$ associated
with $\mathcal{H}$ will be used extensively.

The notations for continuous parametrized curves in $\mathcal{B}$
and $\mathcal{R}$ are as follows:
\begin{eqnarray}
&& \psi_1, \psi_2\in \mathcal{B}: \mathcal{C}=\{\psi(s)|s_1\le s\le s_2\}\subset \mathcal{B},\nonumber\\
&&~~~~~~~~~~~~~~~~~~~~\psi(s_1)=\psi_1, \psi(s_2)=\psi_2;\nonumber\\
&& C=\pi(\mathcal{C})=\{\rho(s)=\psi(s)\psi(s)^\dagger| s_1\le s\le
s_2\}\subset \mathcal{R},\nonumber\\
&& \rho(s_1)=\rho_1, \rho(s_2)=\rho_2.
\end{eqnarray}
Any $\mathcal{C}$ projecting onto a given $C$ is a lift of the
latter. The prerequisites for the geometric phase to be defined for
$C, \mathcal{C}$ are:
\begin{eqnarray}
Tr(\rho_1\rho_2)>0,\quad (\psi_1, \psi_2)\ne 0;\nonumber\\
\rho(s), \psi(s) \quad \hbox{piecewise once differentiable}.
\end{eqnarray}
(Here $C$ need not be closed; and even if it is, $\mathcal{C}$ could
be open). The prerequisites for $C$ to be a NPC are:
\begin{eqnarray}
s', s\in [s_1, s_2]: Tr(\rho(s')\rho(s))> 0;\nonumber\\
\rho(s)\ \hbox{once differentiable for all}\ s\in [s_1, s_2].
\end{eqnarray}

For such $C$ we will use lifts $\mathcal{C}$ obeying similar
conditions:
\begin{eqnarray}
s', s\in [s_1, s_2]: (\psi(s'), \psi(s))\ne 0;\nonumber\\
\psi(s)\ \hbox{once differentiable for all}\ s\in [s_1, s_2].
\end{eqnarray}

\subsection*{Definition of a NPC}

Given conditions (4.3, 4.4) we have two equivalent ways to define a
NPC. $C\subset \mathcal{R}$ is a NPC if either
\begin{eqnarray}
s, s', s''\in [s_1, s_2]: \Delta_3(\psi(s), \psi(s'), \psi(s''))\nonumber\\
=Tr(\rho(s)\rho(s')\rho(s''))=\hbox{real}>0;
\end{eqnarray}
or if for any fixed $s_0\in [s_1, s_2]$,
\begin{equation}
s, s'\in [s_1, s_2]: \Delta_3(\psi(s_0), \psi(s),
\psi(s'))=\hbox{real}>0.
\end{equation}
We denote NPC's in $\mathcal{R}$ by $N, N', \ldots$. Any lift of $N$
will be written as $\mathcal{N}$, and will also be called a NPC (in
$\mathcal{B}$).

\subsection{Properties of NPC's}
For any lift $\mathcal{N}$ of a NPC $N\subset \mathcal{R}$ from
$\rho_1$ to $\rho_2$,
\begin{equation}
-i\int_{\mathcal{N}s_1}^{s_2}{\rm d} s (\psi(s),
\frac{d\psi(s)}{ds})=\arg (\psi_1, \psi_2).
\label{4.7}
\end{equation}
For any NPC $N\subset \mathcal{R}$, there exist lifts
$\mathcal{N}_0\subset \mathcal{B}$ such that
\begin{equation}
s', s\in [s_1, s_2]: (\psi_0(s'), \psi_0(s))=\hbox{real}>0.
\label{4.8}
\end{equation}

\subsection*{Construction of most general NPC}

Let $\rho_1, \rho_2$ be distinct nonorthogonal points in
$\mathcal{R}$. Choose $\psi_1, \psi_2\in \mathcal{B}$ projecting on
to them and obeying the Pancharatnam `in phase' condition $(\ref{2.5})$
which is repeated here:
\begin{equation}
(\psi_1, \psi_2)=\cos\frac{1}{2}\theta_0, \quad 0<\theta_0<\pi.
\end{equation}
Then there exists an ON pair of vectors $\{e_1, e_2\}$ such that
\begin{equation}
\psi_1=e_1, \psi_2=e_1\cos\frac{1}{2}\theta_0+
e_2\sin\frac{1}{2}\theta_0.
\label{4.10}
\end{equation}
The (unique) geodesic from $\psi_1$ to $\psi_2$ is a reparametrised
form of eq. $(\ref{2.6})$:
\begin{eqnarray}
\psi(s)&=&\frac{1}{\sin\frac{1}{2}\theta_0}\{\psi_1 \sin(\frac{1}{2}\theta_0\frac{(s_2-s)}{(s_2-s_1)}) + \psi_2  \sin(\frac{1}{2}\theta_0 \frac{(s-s_1)}{(s_2-s_1)})\}\nonumber\\
&& =e_1\cos(\frac{1}{2}\theta_0\frac{(s-s_1)}{(s_2-s_1)}) + e_2
\sin(\frac{1}{2} \theta_0 \frac{(s-s_1)}{(s_2-s_1)}), \nonumber\\
&& s_1\le s\le s_2.
\end{eqnarray}

The steps to follow to obtain a lift $\mathcal{N}_0$ (of the type
obeying eq. $(\ref{4.8})$) of an NPC $N$ from $\psi_1$ to $\psi_2$ are
these:

(a) Extend the ON pair $\{e_1, e_2\}$ to an ONB $\{e_1, e_2, e_3,
\ldots, e_n\}$ for $\mathcal{H}$ in any way.

(b) Define $\psi_0(s), s\in [s_1, s_2]$, by
\begin{equation}
\psi_0(s) = \sum_{r=1}^{n}x_r(s)e_r
\label{4.12}
\end{equation}
where
\begin{equation}
{\bf x}(s)=(x_1(s), x_2(s), \ldots, x_n(s))
\end{equation}
is a real $n$-component vector obeying three types of conditions:

(c) Boundary conditions:
\begin{equation}
{\bf x}(s_1)=(1, 0, 0, \ldots, 0), {\bf x}(s_2)=(
\cos\frac{1}{2}\theta_0, \sin\frac{1}{2}\theta_0, 0, \ldots, 0).
\end{equation}

(d) Local conditions: for $s\in (s_1, s_2)$,
\begin{eqnarray}
&& {\bf x}(s)\cdot {\bf x}(s)=\sum_{r=1}^n x_r(s)^2=1;\nonumber\\
&& x_1(s), x_1(s) \cos\frac{1}{2}\theta_0 +x_2(s)\sin\frac{1}{2}\theta_0>0;\nonumber\\
&& {\bf x}(s)\  \hbox{continuous once-differentiable}.
\end{eqnarray}
Thus for all $s\in [s_1, s_2],  {\bf x}(s)\in \mathbb{S}^{n-1}$.

(e) Nonlocal conditions:
\begin{equation}
s', s\in [s_1, s_2]: (\psi_0(s'), \psi_0(s))= {\bf x}(s')\cdot {\bf
x}(s)\in (0, 1].
\label{4.16}
\end{equation}
Then
\begin{equation}
\mathcal{N}_0=\{\psi_0(s)|s_1\le s\le s_2\}\subset \mathcal{B}
\label{4.17}
\end{equation}

is a lift (of the type $(\ref{4.8})$) of an NPC $N\subset \mathcal{R}$ from
$\rho_1$ to $\rho_2$; and all possible $N$ will be obtained in this
way.

The most general solution to the nonlocal condition $(\ref{4.16})$ cannot be
easily given explicitly in any way. If we restrict the choice of
${\bf x}(s)$ by the condition $x_r(s)\ge 0, r=2, 3, \ldots, n$,
conditions $(\ref{4.16})$ are immediately obeyed. However, in the most
general case, $x_r(s)$ for some $r\in (2, 3, \ldots n)$ can be
negative for some ranges of $s$.

Step (a) in the above construction shows why nongeodesic NPC's exist
only for $n\ge 3$.

\subsection*{Generalised BI-geometric phase connection}

This is an extension of eq. $(\ref{2.8})$. We can replace each of the
geodesics on the left by any NPC with the same endpoints. Use of the
property $(\ref{4.7})$ of (lifts of) NPC's leads in an obvious notation to
the result
\begin{equation}
\varphi_{geom}[N_{12}\cup N_{23}\cup N_{31}]=-\arg \Delta_3(\psi_1,
\psi_2, \psi_3).
\label{4.18}
\end{equation}

\section{Bargmann invariants and Null Phase Curves in the Schwinger-Majorana SU(2) framework}

\setcounter{equation}{0}

The widest possible generalisation of the original BI-geometric
phase connection $(\ref{2.8})$ is given by eq. $(\ref{4.18})$. On the left hand
side, geodesics have been replaced by the much more plentiful NPC's
(for each given pair of vertices). On the right, the BI stays
unchanged. Here it is of course assumed that $\dim \mathcal{H}=n\ge
3$.

In this Section we apply the ideas of the Schwinger-Majorana
framework for SU(2) to both sides of the connection $(\ref{4.18})$. We first
deal with the right hand side for general $n$, and see how it
reduces in principle to calculations on Poincar\'{e} spheres. The
case $n=3$ is worked out in some detail. Next we study the left hand
side, for $n=3$, to see how the difference between geodesics and
general NPC's appears in the Schwinger-Majorana framework.

\subsection*{Treatment of the BI}

In dealing with vectors in $\mathcal{H}_n^{(Sch)}$, we have the
freedom to use unitary transformations within the UIR $D^{(J)}(u)$
of SU(2), or to use the wider set of transformations in {U}(n). (In
what follows, $n$ is kept fixed). To study the BI $\Delta_3(\psi_1,
\psi_2, \psi_3)$ in general terms, in the spirit of Section II we
exploit U(n) action. However, at the same time we use the
information about the description of vectors from the SU(2) point of
view, given in the Appendix.

Let the triad $|\psi_j\rangle\in \mathcal{H}_n^{(Sch)}, j=1, 2, 3$,
determine the six intrinsic and algebraically independent angles
$\theta_{jk}, \varphi_{jk}$ as in eq. $(\ref{2.10})$. In the present
situation, a preferred ONB $(\ref{A.4})$ in $\mathcal{H}_n^{(Sch)}$ is
already given by SU(2) considerations. With the help of a suitable
U(n) transformation we can always map $|\psi_1\rangle$ to the
highest weight vector $|J, J\rangle$. We indicate this by
\begin{equation}
|\psi_1\rangle\mathop{\longrightarrow}^{U\in \text{U}(n)}
|\psi'_1\rangle=|J, J\rangle.
\label{5.1}
\end{equation}
To preserve the BI this same U(n) transformation must be applied
to $|\psi_2\rangle, |\psi_3\rangle $ as well. To avoid making the
notation excessively intricate, such steps will be left implicit.

The stability group of $|\psi'_1\rangle$ is clearly a U(n-1)
subgroup of U(n). We can next always use this U(n-1) action
(after the U(n) action $(\ref{5.1})$) to map $|\psi_2\rangle$ to some pure
product state $|\xi; n\rangle$:
\begin{eqnarray}
|\psi_2\rangle\mathop{\longrightarrow}^{U(n-1)} |\psi'_2\rangle=|\xi; n\rangle,\nonumber\\
\xi=\left(\begin{array}{c}\alpha\\ \beta\end{array}\right),
\xi^\dagger\xi=1,
\label{5.2}
\end{eqnarray}
provided (see eq. $(\ref{A.15})$)
\begin{equation}
\langle \psi'_1 |\psi'_2\rangle =\langle (1, 0);n|\xi;
n\rangle=\alpha^{n-1}=e^{i\varphi_{12}}\cos\frac{1}{2}\theta_{12}.
\label{5.3}
\end{equation}
This essentially determines $\alpha$ but leaves the phase of $\beta$
undetermined:
\begin{eqnarray}
\alpha=e^{i\varphi_{12}/(n-1)}(\cos\frac{1}{2}\theta_{12})^{1/(n-1)},\nonumber\\
|\beta|=(1-(\cos\frac{1}{2}\theta_{12})^{2/(n-1)})^{1/2}.
\label{5.4}
\end{eqnarray}

Turning finally to $|\psi_3\rangle$, it is clear that in general we
cannot expect to be able to map it to some pure product state
$|\xi'; n\rangle$, because (as shown in Section II) with
two-dimensional systems only five algebraically independent angles
can be accommodated. Thus the best that can be achieved in general,
using U(n) action, is that $|\psi_1\rangle, |\psi_2\rangle$ go
into pure product states while $|\psi_3\rangle$ goes into a general
Majorana state:
\begin{eqnarray}
&& |\psi_j\rangle\mathop{\longrightarrow}^{\text{U}(n)} |\psi'_j\rangle: |\psi'_1\rangle=|J, J\rangle=|(1, 0); n\rangle;\nonumber\\
&& |\psi'_2\rangle=|\xi; n\rangle, \xi=\left(\begin{array}{c}\alpha\\ \beta\end{array}\right);\nonumber\\
&& |\psi'_3\rangle=c|\{\xi'\}; n\rangle, \{\xi'\}=\{\xi'_1, \xi'_2,
\ldots, \xi'_{n-1}\},\ \hbox{unordered},\nonumber\\
\label{5.5}
\end{eqnarray}
where $c$ is a normalising constant:
\begin{equation}
c=|\langle \{\xi'\}; n| \{\xi'\}; n\rangle|^{-1/2}.
\end{equation}
Then, given $\theta_{31}$ and $\varphi_{31}$, use of eq. $(\ref{A.16})$
leads to
\begin{eqnarray}
&&\langle \psi'_3|\psi_1'\rangle= c\sqrt{(n-1)!}\prod_{k=1}^{n-1}\alpha'^*_k=e^{i\varphi_{31}}\cos\frac{1}{2}\theta_{31},\nonumber\\
&& \xi'_k=\left(\begin{array}{c}\alpha'_k\\
\beta'_k\end{array}\right),
\label{5.7}
\end{eqnarray}
giving partial information on $\{\xi'\}$. The phase freedom of
$\beta$ in eq. $(\ref{5.4})$, and in the choice of $\{\xi'\}$ after
requiring eq. $(\ref{5.7})$, are analogues of the presence of independent
angles $\phi$ in eq. $(\ref{3.5})$, $\phi$ and $\xi$ in eq. $(\ref{3.24})$. All
these remaining freedoms while reproducing given $\theta_{jk},
\varphi_{jk}$ must be kept in mind.

Now we turn to the BI which by U(n) invariance becomes
\begin{eqnarray}
&& \Delta_3(\psi_1, \psi_2, \psi_3)= \Delta_3(\psi'_1, \psi'_2, \psi'_3)\nonumber\\
&&=c^2\langle (1, 0); n|\xi; n\rangle\langle \xi;n|\{\xi'\};n\rangle\langle\{\xi'\};n|(1, 0);n\rangle\nonumber\\
&& =c^2(n-1)!\alpha^{n-1}\prod_{k=1}^{n-1}\xi^\dagger \xi'_k\cdot  \prod_{k'=1}^{n-1}\alpha^*_{k'}\nonumber\\
&& = c^2(n-1)!
\prod_{k=1}^{n-1}\Delta_3\left(\left(\begin{array}{c}1\\
0\end{array}\right), \xi, \xi'_k\right).
\end{eqnarray}

The righthand side of eq. $(\ref{4.18})$ then becomes
\begin{eqnarray}
-\arg \Delta_3(\psi_1, \psi_2, \psi_3)=-\sum_{k=1}^{n-1}\arg
\Delta_3\left(\left(\begin{array}{c}1\\ 0\end{array}\right), \xi,
\xi'_k\right),
\label{5.9}
\end{eqnarray}
so the general $n$-level system geometric phase as viewed from the
BI is the sum of $(n-1)$ geometric phases of Pancharatnam type
computed using $(n-1)$ geodesic triangles on the Poincar\'{e}'
sphere \cite{11}.

This result is of a mathematical nature, without implying any
physical substructure for the $n$-level quantum system. It may also
be viewed as a general structure analysis of the $n$-level
three-vertex geometric phase, not as an explicit evaluation of it in
the sense of eqs. $(\ref{3.27})$.

It is instructive to work out the above expressions in more detail
for the lowest nontrivial case $n=3$. The spin 1 UIR of SU(2),
$D^{(1)}(u)$, as well as the nine parameter group U(3), act on
$\mathcal H_3^{(Sch)}$. (The former is equivalent to the real
defining representation of SO(3)). From the SU(2) point of view, we
have the following ONB and vector descriptions:
\begin{eqnarray}
&&\hbox{ONB}\nonumber\\ &&
e_1=\frac{(\hat{a}_1^+)^2}{\sqrt{2}}|0, 0\rangle=|2,
0\rangle=\left|\left(\begin{array}{c}1 \\ 0\end{array}\right); 3\right\rangle,\nonumber\\
&& e_2=\hat{a}_1^+\hat{a}_2^+ |0, 0\rangle=|1,
1\rangle=\left|\left\{\left(\begin{array}{l}1 \\
0\end{array}\right), \left(\begin{array}{l}0 \\
1\end{array}\right)\right\};
3\right\rangle,\nonumber\\
&& e_3=\frac{(\hat{a}_2^+)^2}{\sqrt{2}}|0, 0\rangle=|0,
2\rangle=\left|\left(\begin{array}{l}0 \\ 1\end{array}\right);
3\right\rangle;\hspace{3pc} (a)\nonumber\\
&&\hbox{General (Majorana) vectors}\nonumber\\ && \xi= \left(\begin{array}{l}\alpha \\
\beta\end{array}\right), \xi'= \left(\begin{array}{l}\alpha' \\
\beta'\end{array}\right), \xi^\dag\xi=\xi'^\dag\xi' =1:\nonumber\\
&& |\{\xi, \xi'\}, 3\rangle=|\{\xi', \xi\}; 3\rangle= (\alpha
\hat{a}_1^+ +\beta \hat{a}_2^+) (\alpha' \hat{a}_1^+ +\beta'
\hat{a}_2^+)|0, 0\rangle\nonumber\\
&& =\sqrt{2}\alpha
\alpha'e_1+(\alpha\beta'+\beta\alpha')e_2+\sqrt{2}\beta\beta'e_3; \hspace{2pc} (b)\nonumber\\
&&\hbox{Pure product vectors}\nonumber\\ && \xi= \left(\begin{array}{l}\alpha \\
\beta\end{array}\right), \xi^\dag\xi=1:\nonumber\\
&& |\xi; 3\rangle=\frac{1}{\sqrt{2}}|\{\xi, \xi\}; 3\rangle=
\frac{(\alpha\hat{a}_1^+ +\beta \hat{a}_2^+)^2}{\sqrt{2}}|0,
0\rangle\nonumber\\
&&=\alpha^2e_1+\sqrt{2}\alpha\beta e_2+\beta^2e_3. \hspace{7.5pc} (c)\nonumber\\
\label{5.10}
\end{eqnarray}
Out of the ONB vectors, only $e_1$ and $e_3$ are of pure product
type. The general vector $|\{\xi, \xi'\}, 3\rangle$ determines an
unordered pair of points $\{\hat{n}, \hat{n}'\}$ on the Poincar\'{e}
sphere: \begin{equation}\hat{n}=\xi^\dag\ \underline{\sigma}\ \xi,
\quad \hat{n}'=\xi'^\dag\ \underline{\sigma}\ \xi'.
\end{equation}
For the pure product vector, $\hat{n}'=\hat{n}$. So for the  ONB
vectors we have
\begin{eqnarray}
&&e_1\rightarrow \hat{n}'=\hat{n}= (0, 0, 1); e_2\rightarrow \hat{n}=(0, 0, 1),
\hat{n}'= (0, 0, -1); \nonumber\\
&&e_3\rightarrow \hat{n}'=\hat{n}= (0, 0, -1).
\end{eqnarray}

Now let $\psi_j, j=1, 2, 3$ be a triad of (normalized) vectors in
$\mathcal H_3^{(Sch)}$, with the set of six independent U(3)
invariant angles $\theta_{jk}, \varphi_{jk}$. From eq. $(\ref{5.1})$ we know
that with no loss of generality we can assume
\begin{equation}\psi_1=e_1= \left|\left(\begin{array}{l}1 \\ 0\end{array}\right);
3\right\rangle\rightarrow \hat{n}_1=\hat{n}'_1= (0, 0,
1).
\label{5.13}
\end{equation} From eqs $(\ref{5.2}, \ref{5.3}, \ref{5.4})$ we may next assume
$\psi_2$ is also of pure product type:
\begin{eqnarray}&&\psi_2=|\xi_2; 3\rangle, \xi_2= \left(\begin{array}{l}\alpha_2 \\ \beta_2\end{array}\right)
=e^{i\varphi_{12}/2} \left(\begin{array}{l}\sqrt{C_{12}} \\
\sqrt{1-C_{12}}\end{array}\right)\rightarrow\nonumber\\
&&\hat{n}_2=\hat{n}'_2=(2\sqrt{C_{12}(1-C_{12})}, 0,
2C_{12}-1).
\label{5.14}
\end{eqnarray} Here we have chosen $\arg \beta_2$
conveniently, and then $\theta_{12}, \varphi_{12}$ are properly
incorporated. The choice of a convenient form for $\psi_3$ is more
involved. From eq. ($\ref{5.5})$ we know that in general it is of general
type:
\begin{eqnarray}&&\psi_3=c|\{\xi_3, \xi'_3\}; 3\rangle,
c=(1+|\xi^\dag_3\xi'_3|^2)^{-1/2},\nonumber\\
&&\xi_3= \left(\begin{array}{l}\alpha_3 \\
\beta_3\end{array}\right),
\xi'_3= \left(\begin{array}{l}\alpha'_3 \\
\beta'_3\end{array}\right).
\end{eqnarray} Now on the one hand
$\xi_3, \xi'_3$ are limited by the given values of $\theta_{23},
\theta_{31}, \varphi_{23}, \varphi_{31}$:
\begin{eqnarray}
&& (\psi_2, \psi_3)=e^{i\varphi_{23}}C_{23}=c\sqrt{2}
\xi^\dag_2\xi_3\xi^\dag_2\xi'_3,\nonumber\\
&& (\psi_3, \psi_1)=e^{i\varphi_{31}}C_{31}=c\sqrt{2}
(\alpha_3\alpha'_3)^*.
\label{5.16}
\end{eqnarray}
These conditions determine $\psi_3$, i.e. $\{\xi_3, \xi'_3\}$, upto
a one-parameter group of U(1) transformations in the one-dimensional
subspace of $\mathcal H_3^{(Sch)}$ orthogonal to both $\psi_1$ and
$\psi_2$. However, for all these $\psi_3$'s the geometric phase is
the same, as seen explicitly in eqs. $(\ref{3.1})$. On the other hand they
determine the pair of points on the Poincar\'{e} sphere
corresponding to $\psi_3$:
\begin{equation}
\xi^\dag_3\ \underline{\sigma}\ \xi_3=\hat{n}_3, \quad \xi'^\dag_3\
\underline{\sigma}\ \xi'_3=\hat{n}'_3.
\end{equation}
From eqs $(\ref{5.13}, \ref{5.14})$, both $\hat{n}_1=\hat{n}'_1$ and
$\hat{n}_2=\hat{n}'_2$ lie on the `Greenwich' meridian with
azimuthal angle $\phi=0$. To then allow for the most general choice
of $\psi_3$ we must allow $\hat{n}_3, \hat{n}'_3$ to be any
independently chosen pair of points on the Poincar\'{e} sphere.
Using spherical polar variables we thus have
\begin{equation}
\hat{n}_3\rightarrow (\theta_3, \phi_3), \hat{n}'_3\rightarrow
(\theta'_3, \phi'_3), \quad 0\le \theta_3, \theta'_3\le \pi, 0\le
\phi_3, \phi'_3<2\pi.
\end{equation}
When the two complex equations $(\ref{5.16})$ are viewed as conditions on
$\hat{n}_3, \hat{n}'_3$ the earlier remarks tell us that the four
real angles $\theta_3, \theta'_3, \phi_3, \phi'_3$ are determined by
$\theta_{23}, \theta_{31}, \varphi_{23}, \varphi_{31}$ upto the
freedom of U(1) transformations mentioned above, but this does not
affect the geometric phase. For the calculation of this phase, and
picturing it on the Poincar\'{e} sphere, the former angles are more
convenient. From Pancharatnam's well known result for two-level
systems, eq. $(\ref{5.9})$ becomes
\begin{equation} -\arg\Delta_3(\psi_1, \psi_2,
\psi_3)=\frac{1}{2}\Omega(\hat{n}_1, \hat{n}_2, \hat{n}_3)+
\frac{1}{2}\Omega(\hat{n}_1, \hat{n}_2, \hat{n}'_3)
\end{equation}
where each $\Omega(\ldots)$ is the solid angle of the spherical
triangle with indicated vertices (counted positive if the sequence
of vertices appears anticlockwise when viewed from the outside).

\subsection*{NPC's in Schwinger--Majorana framework -- general structure analysis}

To keep the various expressions as simple as possible, we consider
only the case of $\mathcal{H}_3^{(Sch)}$ corresponding to spin
$J=1$. This is the lowest dimension in which nontrivial NPC's occur.

Given two normalised vectors $\psi_1, \psi_2\in
\mathcal{H}_3^{(Sch)}$ with $(\psi_1,
\psi_2)=\cos\frac{1}{2}\theta_0, \theta_0\in (0, \pi)$, we wish to
describe a general NPC, and contrast it with the geodesic,
connecting them. (The latter is given, in the case of any Hilbert
space $\mathcal{H}$, by eq. $(\ref{2.6})$.) In each case at each point we
wish to find and visualise the corresponding unordered pair
$\{\hat{n}', \hat{n}\}$ of points on $\mathbb{S}^2$. All this will use the
review of NPC's in Section III.

From the pair $\psi_1, \psi_2$ we extract an ON pair written as
$\{e'_1, e'_2\}$ (to be distinguished from $e_1, e_2$ already used
in eq. $(\ref{5.10}(a))$), and have in place of eq. $(\ref{4.10})$
\begin{equation}
\psi_1=e'_1,
\psi_2=e'_1\cos\frac{1}{2}\theta_0+e'_2\sin\frac{1}{2}\theta_0.
\label{5.20}
\end{equation}
From the U(3) point of view there is no intrinsic difference or
distinction between one ON pair and another, as one can be
transformed into the other. However from the SU(2) point of view
there are intrinsically distinct possibilities which cannot be
connected by SU(2) transformations. (As can easily be verified,
SU(2) transformations on $\mathcal{H}_3^{(Sch)}$ take pure product
states into other pure product states, and general (Majorana)
vectors to other such vectors.) Thus each of $e'_1$ and $e'_2$ can
be a pure product vector or a general (Majorana) vector. We now
analyse two of these three cases, to illustrate the kinds of
configurations that can arise.

\subsection*{The pure--pure case}

Suppose both $e'_1$ and $e'_2$ are pure product states:
\begin{equation}
e'_1=|\xi'_1\rangle, e'_2=|\xi'_2\rangle , \xi'^\dagger_2 \xi'_1=0.
\end{equation}
Then both belong to the SU(2) orbit of $e_1$. We can easily see that
by a suitable SU(2) transformation followed by another suitable U(3)
transformation, this pair can be mapped to the pair $e_1, e_3$ of
eq. $(\ref{5.10}$(a)):
\begin{equation}
e'_1, e'_2\rightarrow e_1, e_3.
\label{5.22}
\end{equation}
Let us for convenience use the short hand symbols
\begin{equation}
C_0=\cos\frac{1}{2}\theta_0, S_0=\sin\frac{1}{2}\theta_0,
C=\cos\frac{s}{2}\theta_0, S=\sin\frac{s}{2}\theta_0,
\end{equation}
all of which are nonnegative for $0\le s\le 1$. Then we assume
\begin{equation}
\psi_1=e_1, \psi_2=C_0e_1+S_0e_3.
\label{5.24}
\end{equation}
The geodesic $\mathcal{C}_{geo}$ connecting them is
\begin{eqnarray}
&&\mathcal{C}_{geo}=\{\psi(s)|0\le s\le 1\}:\nonumber\\
&& \psi(s)=Ce_1+Se_3=\frac{1}{\sqrt{2}}(C\hat{a}^{\dagger^2}_1 +S\hat{a}^{\dagger^2}_2)|0, 0\rangle\nonumber\\
&& =\frac{1}{\sqrt{2}} (C^{1/2}\hat{a}^{\dagger}_1 +i
S^{1/2}\hat{a}^{\dagger}_2)
(C^{1/2}\hat{a}^{\dagger}_1 -iS^{1/2}\hat{a}^{\dagger}_2)|0, 0\rangle\nonumber\\
&&=\frac{1}{\sqrt{2}}(C+S)|\xi'(s), \xi(s)\rangle,\nonumber\\
&& \xi'(s)=\frac{1}{\sqrt{C+S}}\left(\begin{array}{c}C^{1/2}\\
iS^{1/2}\end{array} \right),\nonumber\\
&&\xi(s)=\frac{1}{\sqrt{C+S}}\left(\begin{array}{c}C^{1/2}\\-
iS^{1/2}\end{array} \right).
\label{5.25}
\end{eqnarray}

The Majorana pair of vectors $\{\hat{n}'(s), \hat{n}(s)\}$ on $\mathbb{S}^2$
representing $\psi(s)$ along $\mathcal{C}_{geo}$ is then
\begin{eqnarray}
&&\hat{n}'(s)=\frac{1}{(C+S)}(0, 2\sqrt{CS}, C-S),\nonumber\\
&&\hat{n}(s)=\frac{1}{(C+S)}(0, -2\sqrt{CS}, C-S).
\label{5.26}
\end{eqnarray}
Both vectors are on the 2--3 meridian of $\mathbb{S}^2$, reflections of each
other in the 1--3 plane.

The most general NPC $\mathcal{N}_0$ from $\psi_1$ to $\psi_2$,
obeying eq. $(\ref{4.8})$, is constructed following the sequence of steps in
Section III, eqs. $(\ref{4.12}-\ref{4.17})$. The first step is to extend $\{e_1,
e_3\}$ to an ONB for $\mathcal{H}_3^{(Sch)}$ in the most general
way. Thus for any fixed $\eta$, we adjoin $e^{i\eta}e_2$ to  $\{e_1,
e_3\}$. Then the most general $\mathcal{N}_0$ is
\begin{eqnarray}
&& \mathcal{N}_0=\{\psi_0(s)|0\le s\le 1\}:\nonumber\\
&& \psi_0(s)=x_1(s)e_1+x_2(s)e_3+x_3(s)e^{i\eta}e_2,
\label{5.27}
\end{eqnarray}
where ${\bf x}(s)$ obeys:
\begin{eqnarray}
&& {\bf x}(s)\ \hbox{real}, {\bf x}(s)\cdot {\bf x}(s)=1;\nonumber\\
&&{\bf x}(0)=(1, 0, 0), {\bf x}(1)=(C_0, S_0, 0);\nonumber\\
&& x_1(s), C_0x_1(s)+S_0x_2(s)> 0;\nonumber\\
&& 0<{\bf x}(s')\cdot {\bf x}(s)\le 1,
\label{5.28}
\end{eqnarray}
in  addition to being continuous once-differentiable. Compared to
the geodesic $(\ref{5.25})$ where ${\bf x}(s)=(C, S, 0)$, it is $x_3(s)$
that is new. For the Majorana pair along $\mathcal{N}_0$ we need to
factorize the quadratic in $\psi_0(s)$:
\begin{eqnarray}
 \psi_0(s)&=&\frac{1}{\sqrt{2}}\{x_1\hat{a}^{\dagger^2}_1+\sqrt{2}e^{i\eta}x_3\hat{a}^{\dagger}_1\hat{a}^{\dagger}_2+x_2\hat{a}^{\dagger^2}_2\}|0, 0\rangle\nonumber\\
&
=&\frac{x_1}{\sqrt{2}}\left(\hat{a}^{\dagger}_1+\frac{e^{i\eta}x_3+\sqrt{e^{2i\eta}x_3^2-2x_1x_2}}{\sqrt{2}x_1}\hat{a}^{\dagger}_2\right)\nonumber\\
&&\times\left(\hat{a}^{\dagger}_1+\frac{e^{i\eta}x_3-\sqrt{e^{2i\eta}x_3^2-2x_1x_2}}{\sqrt{2}x_1}\hat{a}^{\dagger}_2\right)|0,
0\rangle.\nonumber\\
\end{eqnarray}
Therefore we have the pair (upto real normalisation factors)
\begin{eqnarray}
&&\xi'_0(s)=\left(\begin{array}{c}x_1\\
\frac{e^{i\eta}x_3-\sqrt{e^{2i\eta}x_3^2-2x_1x_2}}{\sqrt{2}}\end{array}\right),\nonumber\\
&&\xi_0(s)=\left(\begin{array}{l}x_1\\
\frac{e^{i\eta}x_3+\sqrt{e^{2i\eta}x_3^2-2x_1x_2}}{\sqrt{2}}\end{array}\right),
\label{5.30}
\end{eqnarray}
leading to corresponding $\hat{n}'_0(s), \hat{n}_0(s)$ on $\mathbb{S}^2$.
These expressions are somewhat complicated, but compared to eqs
$(\ref{5.25}, \ref{5.26})$ for the geodesic some differences show up: the points
$\hat{n}'_0(s), \hat{n}_0(s)$ are generally not on the 2--3
meridian, and not reflections of one another in the 1--3 plane, but
have general positions on $\mathbb{S}^2$.

\subsection*{The pure--general case}

This occurs when in eq. $(\ref{5.20})$, $e'_1$ (say) is a pure product while
$e'_2$ is not:
\begin{eqnarray}
&& e'_1=|\xi'_1\rangle, e'_2=|\xi'_2, \xi''_2\rangle, \nonumber\\
&& \xi'^\dagger_1\xi'_2\ .\ \xi'^\dagger_1\xi''_2=0.
\end{eqnarray}
Now one can see that by a suitable SU(2) transformation $e'_1$ can
be transformed to $e_1$, and therefore at the same time $e'_2$ goes
into some normalised linear combination of $e_2$ and $e_3$. At the
next step by a suitable U(3) transformation preserving $e_1, e'_2$
can be taken to $e_2$. In this way in place of eq. $(\ref{5.22})$ we achieve
\begin{equation}
e'_1, e'_2\rightarrow e_1, e_2,
\end{equation}
and in place of eq. $(\ref{5.24})$ we have the pair
\begin{equation}
\psi_1=e_1, \psi_2=C_0e_1+S_0e_2.
\end{equation}
The geodesic $\mathcal{C}_{geo}$ connecting them is
\begin{eqnarray}
&& \psi(s)=Ce_1+Se_2=\hat{a}^\dagger_1(\frac{C}{\sqrt{2}}\hat{a}^\dagger_1+S\hat{a}^\dagger_2)|0, 0\rangle\nonumber\\
&& =|\xi', \xi(s)\rangle, \nonumber\\
&& \xi'=\left(\begin{array}{c}1\\ 0\end{array}\right), \quad
\xi(s)=\frac{1}{\sqrt{1+S^2}}\left(\begin{array}{c}C\\
\sqrt{2}S\end{array}\right).
\label{5.34}
\end{eqnarray}
In contrast to eq. $(\ref{5.25})$ in the pure-pure case, both of these are
real. The Majorana pair $\{\hat{n}'(s), \hat{n}(s)\}$ is therefore
\begin{equation}
\hat{n}'(s)=(0, 0, 1), \hat{n}(s)=\frac{2}{1+S^2}(\sqrt{2}CS, 0,
\frac{C^2}{2}-S^2).
\label{5.35}
\end{equation}
Here again the contrast with eq. $(\ref{5.26})$ is evident.

For the general NPC from $\psi_1$ to $\psi_2$ in this case we follow
steps similar to the previous pure-pure case. The replacements for
eqs. $(\ref{5.27}-\ref{5.30})$ are:
\begin{eqnarray}
&& \mathcal{N}_0=\{\psi_0(s)|0\le s\le 1\}:\nonumber\\
&& \psi_0(s)=x_1(s)e_1+x_2(s)e_2+x_3(s)e^{i\eta}e_3.
\end{eqnarray}
The conditions on ${\bf x}(s)$ are identical to those given in eq.
$(\ref{5.28})$. Again, compared to the geodesic $(\ref{5.34})$ where  ${\bf
x}(s)=(C, S, 0), x_3$ is new. For the Majorana pair we find:
\begin{eqnarray}
\psi_0(s)&=&\frac{1}{\sqrt{2}}\{x_1\hat{a}^{\dagger^2}_1+\sqrt{2}x_2\hat{a}^{\dagger}_1\hat{a}^{\dagger}_2+x_3e^{i\eta}\hat{a}^{\dagger^2}_2\}|0, 0\rangle\nonumber\\
&
=&\frac{x_1}{\sqrt{2}}\left(\hat{a}^{\dagger}_1+\frac{x_2+\sqrt{x_2^2-2e^{i\eta}x_1x_3}}{\sqrt{2}x_1}\hat{a}^{\dagger}_2\right)\nonumber\\
&\times &\left(\hat{a}^{\dagger}_1+\frac{x_2-
\sqrt{x_2^2-2e^{i\eta}x_1x_3}}{\sqrt{2}x_1}\hat{a}^{\dagger}_2\right)|0,
0\rangle,
\end{eqnarray}
leading to the pair (upto normalisation)
\begin{equation}
\xi'_0(s)=\left(\begin{array}{c}x_1\\ \frac{x_2-
\sqrt{x_2^2-2e^{i\eta}x_1x_3}}{\sqrt{2}}\end{array}\right),
\xi_0(s)=\left(\begin{array}{c}x_1\\
\frac{x_2+\sqrt{x_2^2-2e^{i\eta}x_1x_3}}{\sqrt{2}}\end{array}\right).
\label{5.38}
\end{equation}
This is to be compared on the one hand to the geodesic pair $(\ref{5.34})$,
and on the other hand to the pure-pure NPC case $(\ref{5.30})$. In the pair
$\{\hat{n}'_0(s), \hat{n}_0(s)\}$ on $\mathbb{S}^2$ that follow from eq.
$(\ref{5.38})$: unlike $\hat{n}'(s)$ in eq. $(\ref{5.35})$, $\hat{n}'_0(s)$ is not
constant; and in detailed structure the present pair is different
from that given by eq. $(\ref{5.30})$.

To sum up, in this Section we have analysed the structures of the BI
$\Delta_3(\psi_1, \psi_2, \psi_3)$ and of different kinds of NPC's
using the Schwinger--Majorana SU(2) framework. For any practical
calculation of geometric phases along these lines, specific details
will have to be worked out, but not involving any new points of
principle.

In references 5, the reported values of geometric phases involve
measuring directly phases of BI's, not dealing with continuum
Schr\"odinger evolution with some Hamiltonian along closed paths made
up of geodesics in any state space. From our perspective in this and
the previous Section, it would be interesting to design experiments
involving continuous Schr\"odinger evolution along the sides of a
`triangle' in ray space, in which two sides (say) are geodesics
while the third is a nontrivial (but as simple as possible) NPC.

\section{Concluding Remarks}
The present work focusses on the various objects that appear in the equation 
\begin{equation}
 -\arg \Delta_3(\psi_1,\psi_2, \psi_3)= \varphi_{geom}[N_{12}\cup N_{23}\cup N_{31}]; 
\end{equation}
relating the Pancharatnam phase and the geometric phase and invesigates their structure, properties, convenient parametrizations and 
useful decompositions. 

We show that the third order BI $\Delta_3(\psi_1,\psi_2, \psi_3)$ can be parametrised in terms of six unitary invariant angles. The algebraic 
independence or otherwise of these parameters depends on  the dimension $n$ of the Hilbert space to which the vectors 
$\psi_1,\psi_2, \psi_3$ belong. With no specific assumptions regarding $\psi_1,\psi_2, \psi_3$ ( beyond those stipulated earlier) we show 
that for $n=2$ only five of these angle parameters are algebraically independent. On other hand for $n>2$ all six are algebraically independent. 
This difference in the two cases can be traced back to the fact that in a two dimensional Hilbert space  at most two vectors can be linearly 
independent. As a curiosity we discuss in some detail the case $\psi_1,\psi_2, \psi_3$ are drawn from the set of coherent states of a one 
dimensional harmonic oscillator and find that, like the $n=2$ case, only five unitary invariant angles turn out to be algebraically 
independent although the underlying Hilbert space is infinite dimensional. This unexpected reduction may perhaps be traced back to 
interrelations implied by the overcompleteness of the set of coherent states. For $n=3$ we derive an explicit formula for the third 
order Bargmann invariant in terms of the intrinsic unitary invariant parameters. This, in turn, extends Pancharatnam's result for the 
geometric phase $\varphi_{geom}$ pertaining to a spherical triangle on $\mathbb{S}^2$ to the corresponding case for $n=3$.  

We give an explicit construction for the lifts of the null phase curves $N$ appearing in the BI-geometric phase connection above. 
Combining the ideas presented in ref. [5] on  the use of Majorana representation \cite{6} for symmetric quantum states with 
Schwinger's work \cite{7} on quantum theory of angular momentum,  we develop what we call the Schwinger-Majorana framework for describing states of 
an $n$ level system in which the SU(2) group plays a key role.  We use this framework to develop elegant and convenient descriptions 
for the third order BI and the lifts of the null phase curves and recover the results in ref. [5] expressing the general $n$ level system 
geometric phase as a sum of $(n-1)$ geometric phases of Pancharatnam type computed using $(n-1)$ geodesic triangles on the Poincar\'e sphere.

The three vertex Bargmann invariant plays an important role in quantum information theory in the context of distinguishability of three quantum
 states \cite{12},\cite{13},\cite{14} and we expect that their  description in the Schwinger-Majorana framework as developed here will find useful applications in this area. 
We further hope that our  work will stimulate experimental activity in designing new experiments on geometric phases in higher dimensional systems 
based on direct measurement of the phase of the three vertex Bargmann invariant \cite{5},\cite{15} as well those  involving  evolution along selected
NPC's generated by suitable Hamiltonians.

\begin{acknowledgments}
KSA thanks the University Grant Commission for providing BSR-RFSMS fellowship. NM thanks the Indian National Science
Academy for enabling this work through the INSA
Distinguished Professorship.
\end{acknowledgments}
\renewcommand{\theequation}{\thesection.\arabic{equation}}
\appendix
\section{The Schwinger--Majorana framework for SU(2)}
\setcounter{equation}{0}

The three-parameter group SU(2) is the only compact simple Lie group
which has one unitary irreducible representation (UIR), upto unitary
equivalence, in every finite dimension $n=1, 2, 3, \ldots$. These
UIR's are labelled by the spin or angular momentum quantum number
$J=0, 1/2, 1, \ldots$, with $n=2J+1$. The regular representations of
SU(2) contain each UIR $J$ as often as its dimension $(2J+1)$. A
much `leaner' and very useful unitary representation of SU(2),the
Schwinger representation arising from the Schwinger oscillator
operator construction of the SU(2) Lie algebra \cite{7}, has the
attractive feature that it is the direct sum of all the UIR's of
SU(2), each occurring exactly once. This leads to a specific `model'
of finite dimensional Hilbert spaces $\mathcal{H}_n$ of all
dimensions $n=1, 2, 3, \ldots$, once each, with certain common
operator and vector features arising from SU(2) representation
theory.

It has been recently pointed out that, thanks to the Majorana
theorem for symmetric multispinor UIR's of SU(2), there is a very
natural framework to discuss geometric phases for quantum systems in
any finite dimension \cite{5}. We combine the Schwinger and Majorana
ideas in this Appendix and recall the main features which are used
in Section IV of the main text.

\subsection*{The Schwinger construction}

This is based on two independent quantum mechanical oscillators with
operators obeying the canonical commutation relations
\begin{equation}
[\hat{a}_\alpha, \hat{a}_\beta^\dagger]=\delta_{\alpha\beta},
[\hat{a}_\alpha, \hat{a}_\beta]=0, \alpha, \beta=1, 2.
\label{A.1}
\end{equation}
The hermitian SU(2) generators $\hat{J}_j$ and their commutation
relations are
\begin{eqnarray}
&& \hat{J}_j=\frac{1}{2}\hat{a}_\alpha^\dagger(\sigma_j)_{\alpha \beta}\hat{a}_\beta, j=1, 2, 3;\nonumber\\
&& [\hat{J}_j, \hat{J}_k]=i\epsilon_{jkl}\hat{J}_l.
\end{eqnarray}

The infinite dimensional Hilbert space $\mathcal{H}^{(Sch)}$
carrying an irreducible representation of eqs $(\ref{A.1})$ is the direct
sum of finite dimensional subspaces $\mathcal{H}^{(Sch)}_n$, one for
each $n=1, 2, 3, \ldots$ and mutually orthogonal:
\begin{equation}
\mathcal{H}^{(Sch)}=\sum_{n=1, 2, \ldots}
\oplus\mathcal{H}^{(Sch)}_n.
\end{equation}

The subspace $\mathcal{H}^{(Sch)}_n$ carries the spin
$J=\frac{1}{2}(n-1)$ UIR of SU(2), and can be used as a `model' for
an $n$-level quantum system. An ONB for it is given by
\begin{eqnarray}
&& |J, M\rangle= \frac{(\hat{a}_1^\dagger)^{J+M}(\hat{a}_2^\dagger)^{J-M}}{\sqrt{(J+M)!(J-M)!}}|0, 0\rangle, M=J, J-1,\ldots, -J;\nonumber\\
&& \hat{a}_\alpha|0, 0\rangle=0.
\label{A.4}
\end{eqnarray}
When convenient,  $|J, M\rangle$ will be written as $|n_1,
n_2\rangle$ with $n_1=J+M, n_2=J-M$,  both integral:
\begin{equation}
J=\frac{1}{2}(n_1+n_2), M=\frac{1}{2}(n_1-n_2)=n_1-\frac{1}{2}(n-1).
\end{equation}

A general element $u\in SU(2)$ is unitarily represented on
$\mathcal{H}^{(Sch)}$ as the exponential of $i$ times a real linear
combination of $J_j$:
\begin{equation}
u\in SU(2)\rightarrow D(u)=\exp(-i \alpha_j J_j), |\pmb{\alpha}|\le
2\pi.
\end{equation}
Here $\pmb{\alpha}$ are axis angle parameters; alternatively, using
Euler angles, $D(u)$ is a product of three exponentials. On
$\mathcal{H}^{(Sch)}_n, D(u)$ reduces to the spin $J$ UIR
$D^{(J)}(u)$ of SU(2). Of course, on a given $\mathcal{H}^{(Sch)}_n$
there is also action by $U(n)$ in its defining representation,
containing $D^{(J)}(u)$.

\subsection*{The Majorana theorem, general vectors in $\mathcal{H}^{(Sch)}_n$}

A general vector $|\psi\rangle $ in $\mathcal{H}^{(Sch)}_n$ is
expressible in the ONB $(\ref{A.4})$ as a sum:
\begin{eqnarray}
&& |\psi\rangle=\sum_{M=-J}^J  C_M\frac{(\hat{a}_1^\dagger)^{J+M}(\hat{a}_2^\dagger)^{J-M}}{\sqrt{(J+M)!(J-M)!}}|0, 0\rangle, \nonumber\\
&& \langle\psi|\psi\rangle=\sum_{M=-J}^J |C_M|^2.
\label{A.7}
\end{eqnarray}

Majorana's theorem \cite{6} is essentially the statement that
$|\psi\rangle$ can also be expressed as the product of $(n-1)$
factors each linear in $\hat{a}_\alpha^\dagger$ acting on $|0,
0\rangle$, apart from a constant factor:
\begin{eqnarray}
&& |\psi\rangle=c\prod_{k=1}^{n-1}(\alpha_k \hat{a}_1^\dagger +\beta_k \hat{a}_2^\dagger )|0, 0\rangle, \nonumber\\
&& |\alpha_k|^2+|\beta_k|^2=1, k=1, 2,\ldots, n-1.
\label{A.8}
\end{eqnarray}

If some $\alpha_k$ vanish, the maximum of $M$ in $(\ref{A.7})$ is less than
$J$; while if some $\beta_k$ vanish, the minimum of $M$ exceeds
$-J$. For each $k$ we combine $\alpha_k$ and $\beta_k$ into a
two-component complex column vector $\xi_k$ in a two-dimensional
Hilbert space $\mathcal{H}_2$:
\begin{equation}
\xi_k=\left(\begin{array}{c}\alpha_k\\ \beta_k\end{array}\right),
\xi_k^\dagger\xi_k=1.
\end{equation}
Denote the collection $\xi_1, \xi_2, \ldots, \xi_{n-1}$ by
$\{\xi\}$. Then define the (unnormalised) vectors
\begin{equation}
|\{\xi\}; n\rangle = \prod_{k=1}^{n-1} (\alpha_k\hat{a}_1^\dagger+
\beta_k \hat{a}_2^\dagger) |0, 0\rangle\in \mathcal{H}^{(Sch)}_n,
\end{equation}
with inner products
\begin{equation}
\langle\{\xi'\}; n|\{\xi\}; n\rangle = \sum_{p\in
S_{n-1}}\prod_{k=1}^{n-1} (\xi'_k, \xi_{p(k)}).\end{equation}

From the use of the Poincar\'{e}--Bloch sphere $\mathbb{S}^2$ in
polarization-spin problems, we know that each $\xi$ determines a
point $\hat{n}\in \mathbb{S}^2$, while $\hat{n}$ determines $\xi$ upto a
phase:
\begin{eqnarray}
&& \xi=\left(\begin{array}{c}\alpha\\ \beta\end{array}\right), \xi^\dagger\xi=1:\nonumber\\
&& \hat{n}=\xi^\dagger\ \pmb{\sigma}\ \xi=(2{\rm Re}\ \alpha^*\beta,
2 {\rm Im}\ \alpha^*\beta, |\alpha|^2-|\beta|^2)\in \mathbb{S}^2.\nonumber\\
\end{eqnarray}
If we define $z=\beta/\alpha$, we have the expressions
\begin{eqnarray}
&& \hat{n}=\frac{1}{1+|z|^2}(2{\rm Re}\ z, 2{\rm Im}\ z, 1-|z|^2);\nonumber\\
&& z=\frac{n_1+in_2}{1+n_3}=\frac{1-n_3}{n_1-in_2}.
\end{eqnarray}
As $|z|\rightarrow 0$ or  $\infty, \hat{n}\rightarrow (0, 0, 1)$ or
$(0, 0, -1)$.

Returning to vectors $|\psi\rangle \in \mathcal{H}^{(Sch)}_n$, each
factor in eq. $(\ref{A.8})$ leads to one point $\hat{n}_k\in \mathbb{S}^2$. Since the
operator factors commute, these points are \emph{unordered}. Thus
each $|\psi\rangle$ leads to an unordered set $\{\hat{n}_1,
\hat{n}_2, \ldots, \hat{n}_{2J}\}$ of $(n-1)$ points on $\mathbb{S}^2$;
conversely the latter determines $|\psi\rangle$ upto a complex
factor since
\begin{equation}
\alpha\hat{a}_1^\dagger+\beta\hat{a}_2^\dagger=\alpha(\hat{a}_1^\dagger+\frac{1-n_3}{n_1-in_2}\hat{a}_2^\dagger).
\end{equation}

\subsection*{Pure product vectors in $\mathcal{H}^{(Sch)}_n$}

These are a subset of vectors in $\mathcal{H}^{(Sch)}_n$ which have
a special property with respect to SU(2). They arise when in the
unordered set $\{\xi\}=\{\xi_1, \xi_2, \ldots, \xi_{n-1}\}$, all the
entries are the same. We introduce a simple notation for these
vectors and can easily normalise them:
\begin{eqnarray}
&& \xi_k=\xi=\left(\begin{array}{c}\alpha\\ \beta\end{array}\right), |\alpha|^2+|\beta|^2=1, k=1, 2, \ldots, n-1:\nonumber\\
&& |\xi;n\rangle=\frac{1}{\sqrt{(n-1)!}}|\{\xi, \xi, \ldots,
\xi\};n\rangle=\frac{(\alpha\hat{a}^\dagger_1+\beta
\hat{a}^\dagger_2)^{n-1}}{\sqrt{(n-1)!}}|0, 0\rangle;\nonumber\\
&& \langle J, M|\xi; n\rangle = \sqrt{\frac{(2J)!}{(J+M)!(J-M)!}}\alpha^{J+M} \beta^{J-M};\nonumber\\
&& \langle \xi'; n|\xi;n\rangle =(\xi'^\dagger\xi)^{n-1}.
\label{A.15}
\end{eqnarray}
Another easily obtained inner product is
\begin{equation}
\langle \{\xi'\}; n|\xi;n\rangle =
\sqrt{(n-1)!}\prod_{k=1}^{n-1}(\xi'_k, \xi).
\label{A.16}
\end{equation}

The special SU(2) related property of these vectors is that they are
of highest weight, `$M=J$':
\begin{equation}
\xi^\dagger\ \pmb{\sigma}\ \xi\cdot \hat{\bf J}|\xi; n\rangle=J|\xi;
n\rangle.
\end{equation}
Thus they are SU(2) transforms of $|J, J\rangle$, hence an orbit
under SU(2) action via the spin $J$ UIR in $\mathcal{H}^{(Sch)}_n $.

In summary we have three important results at the vector level in
$\mathcal{H}^{(Sch)}_n $: the ONB $\{|J, M\rangle\}$; the
representation of any $|\psi\rangle$ as a multiple of some
$|\{\xi\}; n\rangle$ for an unordered $\{\xi\}$; and the highest
weight pure product states $|\xi; n\rangle$.



\begin{thebibliography}{99}
\bibitem{11} This is contained in the first of ref. 5.
\end{thebibliography}
\end{document}